\documentclass{aa}  

\usepackage{graphicx}	
\usepackage{txfonts}
\usepackage{hyperref}
\usepackage{placeins}
\usepackage{xcolor}
\usepackage{orcidlink}

\newcommand{\chandra}{Chandra}

\defcitealias{XRISM_M87I}{Paper I}

\begin{document}

   \title{Dynamics of AGN feedback in the X-ray bright East and Southwest `arms' of M87, mapped by XRISM}
   \subtitle{}
   \titlerunning{Dynamics in the X-ray `arms' of M87}
   \authorrunning{The XRISM M87 Target Team}

    \author{A. Simionescu \inst{1,2,3}\fnmsep\thanks{Email: a.simionescu@sron.nl}\orcidlink{0000-0002-9714-3862}
    \and C. Kilbourne \inst{4} \orcidlink{0000-0001-9464-4103}
    \and H.~R. Russell \inst{5} \orcidlink{0000-0001-5208-649X}
    \and D. Ito \inst{6} \orcidlink{0009-0000-4742-5098}
    \and M. Charbonneau \inst{7} 
    \and D. Eckert \inst{8} \orcidlink{0000-0001-7917-3892}
    \and M. Loewenstein \inst{9,4,10} \orcidlink{0000-0002-1661-4029}
    \and J. Martin \inst{2,1}
    \and H. McCall \inst{11} \orcidlink{0000-0003-3537-3491}
    \and B.~R. McNamara \inst{7}
    \and K. Nakazawa \inst{6} \orcidlink{0000-0003-2930-350X}
    \and A. Ogorzalek \inst{9,4,10} \orcidlink{0000-0003-4504-2557}
    \and A. T\"umer \inst{12,4,10} \orcidlink{0000-0002-3132-8776}
    \and I. Zhuravleva \inst{11} \orcidlink{0000-0001-7630-8085}    
    \and N. Dizdar \inst{7} \orcidlink{0009-0003-9080-6736}
    \and Y. Ezoe \inst{13}
    \and R. Fujimoto \inst{14} \orcidlink{0000-0002-2374-7073}
    \and L. Gu \inst{1} \orcidlink{0000-0001-9911-7038}
    \and E. Hodges-Kluck \inst{4} \orcidlink{0000-0002-2397-206X}
    \and Y. Ichinohe \inst{15} \orcidlink{0000-0002-6102-1441}
    \and S. Kitamoto \inst{16} \orcidlink{0000-0001-8948-7983}
    \and M. A. Leutenegger \inst{4} \orcidlink{0000-0002-3331-7595}
    \and F. Mernier \inst{17,9,4,10} \orcidlink{0000-0002-7031-4772}
    \and E. D. Miller \inst{18} \orcidlink{0000-0002-3031-2326}
    \and I. Mitsuishi \inst{6} \orcidlink{0000-0002-9901-233X}
    \and K. Sato \inst{19} \orcidlink{0000-0001-5774-1633}
    \and A. Szymkowiak \inst{20} \orcidlink{0000-0002-4974-687X}
          }

     \institute{
             SRON, Space Research Organisation Netherlands, Niels Bohrweg 4, 2333 CA Leiden, The Netherlands 
        \and Leiden Observatory, Leiden University, PO Box 9513, 2300 RA Leiden, The Netherlands
        \and Kavli Institute for the Physics and Mathematics of the Universe (WPI), The University of Tokyo, Kashiwa, Chiba 277-8583, Japan
        \and NASA / Goddard Space Flight Center, Greenbelt, MD 20771, USA
        \and School of Physics \& Astronomy, University of Nottingham, Nottingham, NG7 2RD, UK
        \and Department of Physics, Nagoya University, Aichi 464-8602, Japan
        \and Department of Physics \& Astronomy, Waterloo Centre for Astrophysics, University of Waterloo, Ontario N2L 3G1, Canada
        \and Department of Astronomy, University of Geneva, Versoix CH-1290, Switzerland
        \and Department of Astronomy, University of Maryland, College Park, MD 20742, USA
        \and Center for Research and Exploration in Space Science and Technology, NASA / GSFC (CRESST II), Greenbelt, MD 20771, USA
        \and Department of Astronomy and Astrophysics, University of Chicago, Chicago, IL 60637, USA
        \and Center for Space Sciences and Technology, University of Maryland, Baltimore County (UMBC), Baltimore, MD, 21250 USA
        \and Department of Physics, Tokyo Metropolitan University, Tokyo 192-0397, Japan
        \and Institute of Space and Astronautical Science (ISAS), Japan Aerospace Exploration Agency (JAXA), Kanagawa 252-5210, Japan
         \and RIKEN Nishina Center, Saitama 351-0198, Japan
        \and Department of Physics, Rikkyo University, Tokyo 171-8501, Japan
        \and IRAP, CNRS, Universit\'{e} de Toulouse, CNES, UT3-UPS, Toulouse, France
        \and Kavli Institute for Astrophysics and Space Research, Massachusetts Institute of Technology, MA 02139, USA
        \and Department of Astrophysics and Atmospheric Sciences, Kyoto Sangyo University, Kyoto 603-8555, Japan
        \and Yale Center for Astronomy and Astrophysics, Yale University, CT 06520-8121, USA
        }

   \date{Received XXX / Accepted XXX}
 
  \abstract
{AGN feedback plays a critical role in regulating gas cooling and star formation in massive galaxies and at the centres of galaxy clusters. As the central galaxy in the nearest cluster, M87 provides the best spatial resolution for disentangling the complex interactions between AGN jets and the surrounding environment.}
{We investigate the velocity structure of the multitemperature X-ray gas in M87, particularly in the eastern and southwestern arms associated with past AGN outbursts, using high-resolution spectroscopy from XRISM/Resolve.}
{We analyze a mosaic of XRISM/Resolve observations covering the core of M87, fitting single- and multi-temperature models to spectra extracted from different regions and energy bands. We assess the line-of-sight velocities and velocity dispersions of the hotter ambient and cooler uplifted gas phases, and evaluate systematic uncertainties related to instrumental gain calibration.}
{The hotter ICM phase, traced by Fe He-$\alpha$ emission, shows velocity dispersions below $\sim100$~km/s, and no significant velocity shifts between the arms and a relaxed offset region, suggesting limited dynamical impact from older AGN lobes. In contrast, the cooler gas phase appears to exhibit larger line of sight velocity gradients up to several hundred km/s as well as a higher velocity dispersion than the ambient hot phase, although these conclusions remain tentative pending improvements in the robustness of the gain calibration at lower energies.}
{The first microcalorimeter-resolved map of gas dynamics in M87 supports the uplift scenario for the X-ray arms, with the cooler gas in the east and southwest seemingly moving in opposite directions along the line of sight. The kinetic energy is a small fraction of the gravitational potential energy associated with the gas uplift, and XRISM further suggests that AGN-driven motions may be short-lived in the hot ambient ICM. These constraints provide important input towards shaping future models of AGN feedback.}

\maketitle

\section{Introduction}

The hot intracluster medium (ICM) in cool-core galaxy clusters exhibits a complex interplay of radiative cooling, active galactic nucleus (AGN) feedback, and minor mergers. High-spatial resolution X-ray observations have revealed a wealth of substructures -- such as cavities, shocks, and filaments -- that directly trace the impact of central AGN outbursts on their surroundings (see reviews by e.g. \citealt{mcnamara2007,hlavacek2022}, and references therein). Among the best-studied examples is Virgo, the nearest galaxy cluster. Its central galaxy, M87, at a distance of only 16.1 Mpc \citep{tonry2001}, hosts a powerful AGN that has repeatedly inflated radio lobes, driving shocks and uplifting cool, metal-rich gas into the ICM \citep[e.g.][]{forman2005,forman2007,million2010,werner2010}. These processes, which play a critical role in heating and redistributing the gas, are expected to imprint characteristic velocity fields in the X-ray emitting plasma. Precise spatial mapping of these velocities will reveal bulk motions due to uplift along the jet axis and turbulence induced by the expansion and propagation of the radio bubbles. Turbulence is thought to be one of the dominant mechanisms dissipating the energy injected by the jets \citep[e.g.][]{zhuravleva2014}, but this can only be conclusively demonstrated with detailed mapping of the gas motions. Constraints on the hot gas dynamics are therefore the crucial missing ingredient limiting our understanding of AGN feedback \citep[see also][for a review]{simionescu2019}.

The launch of the X-Ray Imaging and Spectroscopy Mission \citep[XRISM][]{tashiro2021} has now opened a new window onto the study of cluster gas dynamics. Equipped with the Resolve microcalorimeter, XRISM provides non-dispersive spectroscopy with an energy resolution of $<$5 eV across the 1.7--12 keV band, enabling precise measurements of line centroids and widths even in extended sources \citep[for some examples, see][]{XRISM2025_Centaurus,XRISM2025_Coma,XRISM2025_A2029}. 

In this work, we present XRISM/Resolve observations of M87, focusing on the prominent ``X-ray arms''. These striking, X-ray bright substructures were identified already using Einstein Observatory data \citep{feigelson1987}. They extend to the East and Southwest from the centre of the galaxy, and correlate spatially with the radio emission. Using ROSAT, \cite{boehringer1995} found that the X-ray arms are due to cooler gas, possibly uplifted from the centre of the galaxy. Together with subsequent high-quality radio data by \cite{owen2000}, this led \cite{churazov2001} to argue that the X-ray and radio morphology can be explained by bubbles of radio-emitting plasma rising buoyantly through the hot ICM and entraining thermal gas in their wake. 

Deep $Chandra$ observations further allowed \cite{forman2007} to resolve a web of filamentary structure in both arms. Both arms show narrow filaments in the soft X-ray band, with a projected length-to-width ratio of up to $\sim50$. However, while the Southwestern arm has only a single set of filaments,
the Eastern arm shows a more complex morphology with multiple overlapping sets of filaments and bubbles. Meanwhile, the hard X-ray band (above $\sim$2~keV) shows a much higher degree of azimuthal symmetry, and the arms are not apparent. Instead, the main features seen in this band are two concentric shocks driven by the AGN \citep{forman2007,million2010}. 

Using XMM-Newton data, \cite{simionescu2008} found that the cooler gas phase contained in the X-ray arms is more metal rich than the hotter ambient azimuthally symmetric gas, supporting the scenario where this gas has been uplifted from the vicinity of the central galaxy, where it is enriched by the stellar population of M87. By removing low-entropy gas from the central ICM, that would otherwise cool and form stars, gas uplift is thus an eﬃcient mechanism through which the AGN can both regulate star formation and disperse the metals produced by stars out to larger radii. 

The X-ray arms span radii up to 30~kpc (6--7 arcmin) from the AGN, comparable to the extent of the galaxy itself, and can hence be easily spatially resolved with XRISM’s $\sim 1.3^\prime$ point spread function (PSF). XRISM enables measurements of the line-of-sight velocities and velocity dispersions across the uplifted filaments in M87, allowing us to probe the dynamical state of the uplifted plasma, and its interaction with the surrounding medium. The results presented here provide the first map of ICM gas dynamics in M87 with microcalorimeter precision, offering new insights into the feedback cycle in cool-core clusters.

\section{Observations and data reduction}

\subsection{Observation summary}

The core of M87 was mapped using four distinct XRISM/Resolve aimpoints: one targeting the galaxy's center, two covering the eastern and southwestern radio lobes and corresponding bright soft X-ray `arms', and one located towards the northwest of the core, at a similar radial offset as the arms but away from the regions of evident AGN-ICM interaction. The locations of these pointings are shown in the right panel of Fig. \ref{fig:chandramap}, overlaid on a \chandra\ residual map of M87 which highlights the substructures related to AGN feedback. The NW observation was temporarily paused to carry out a calibration measurement; the SW pointing was interrupted by a Target of Opportunity, and could only be resumed during the following XRISM visibility window of M87, 6 months later. Hence, a total of 6 ObsIDs is available for M87. These are summarised in Table \ref{tab:obs}. 

\begin{table*}[]
    \caption{Details of the XRISM observations used in this work}
    \centering
    \begin{tabular}{ccccccc}
    Pointing & RA & Dec & ObsID & Date & Net exposure & Heliocentric\\
    & & & & & (ks) & correction (km/s) \\
    \hline
      C   & 187.70449 & 12.38992 & 300014010 & 2024-06-13 & 117 & -28.1 \\
      E   & 187.75884 & 12.39092 & 300015010 & 2024-05-30 & 160 & -26.4 \\
      NW1 & 187.67835 & 12.43539 & 300016010 & 2024-05-26 & 26 & -25.3 \\
      NW2 & 187.67858 & 12.43514 & 300016020 & 2024-05-27 & 143 & -25.8 \\
      SW1 & 187.66104 & 12.35739 & 300017010 & 2024-06-02 & 76 & -26.8 \\
      SW2 & 187.66044 & 12.35797 & 300017020 & 2024-12-09 & 82 & 28.4
    \end{tabular}
    \label{tab:obs}
\end{table*}

\subsection{Data reduction and filtering}

We analysed the Resolve data using the XRISM tools publicly available in HeaSOFT v6.34, and version 10 of the calibration database (CalDB). We first reprocessed the data using the task \texttt{xapipeline}, to ensure that the latest calibration is applied to each event file. We then apply the recommended screening criteria based on the \texttt{RISE\_TIME}, \texttt{DERIV\_MAX}, and \texttt{STATUS} flags, as described in the XRISM ABC Guide v1.0 \footnote{\url{https://heasarc.gsfc.nasa.gov/docs/xrism/analysis/abc_guide/xrism_abc.html}}. Pixel 27 is known to exhibit erratic gain jumps which cannot be calibrated with the fiducial cadence of the Fe55 filter wheel data, and was removed from all further analysis. In addition, a similar gain jump was identified from pixel 11 in ObsID 300016020 and therefore this pixel was removed from the spectra of this observation specifically. 

\subsection{Extraction of spectra and response files}
Spectra were accumulated using the \texttt{extractor} task, using only high resolution primary events (Hp). Corresponding response matrices were computed using the tool \texttt{rslmkrmf}. For all spectral analysis, we use the extra-large \texttt{rmfs}, which account most accurately for the secondary response components such as electron-loss continuum and fluorescent escape peaks \citep{eckart2018}. The normalization of the \texttt{rmfs} includes a correction for the branching ratios, in order to recover the correct total flux if only a subset of event grades are selected. It was, however, realised that the presence of many false Ls events (not triggered by real X-ray photons) skews the \texttt{rmf} normalisation to unphysically low values. Therefore, as is now recommended practice for all XRISM/Resolve data, an event file was created from which Ls events were excluded, and this was used as input to \texttt{rslmkrmf}.

\begin{figure*}
    \centering
    \includegraphics[width=\textwidth]{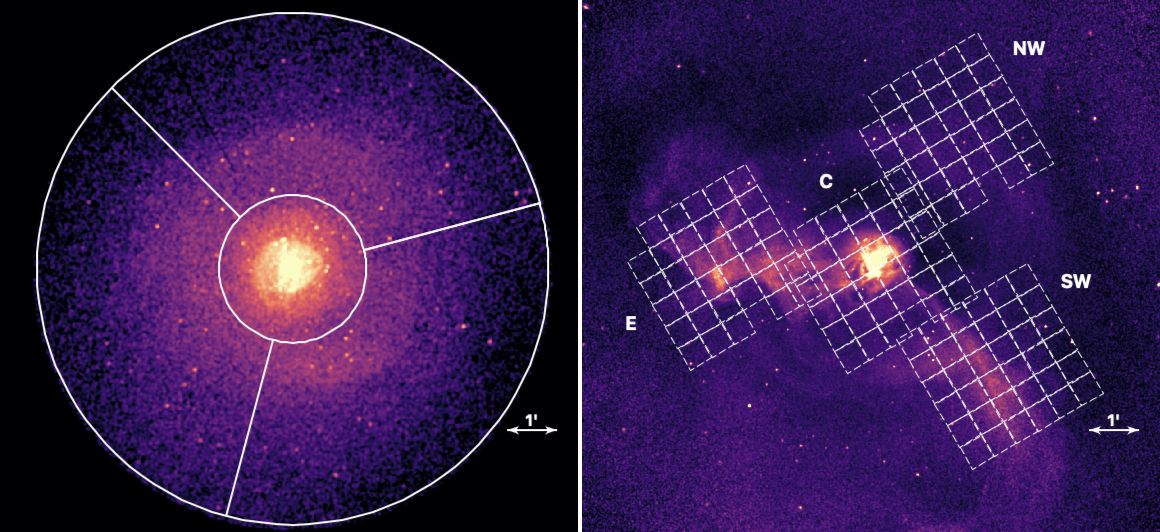}
    \caption{Left: \chandra\ image of M87 in the 3-7 keV energy band. The sky regions used for ray tracing and arf generation are overplotted in white. We neglect scattering from regions located beyond 5.2 arcmin from the center (shown as the outer white circle), hence the image is set to zero here. Right: \chandra\ image of M87 in the 0.5-2 keV energy band, divided by a spherically symmetric beta model to highlight substructure related to AGN feedback. In particular the bright E and SW `arms' can be easily seen. The four XRISM/Resolve fields of view used to map the core of M87 are shown in white. Both images correspond to the same sky coordinates.}
    \label{fig:chandramap}
\end{figure*}

Ancillary response files were calculated using the task \texttt{xaarfgen}. This includes a calculation of raytracing via \texttt{xrtraytrace}, designed to predict the fraction of light that is scattered from any given sky region to any part of the detector, due to XRISM's broad point spread function (PSF), given an input spatial distribution of the source.   
To model the thermal emission from the ICM, we computed \texttt{arfs} using the IMAGE mode of \texttt{xaarfgen}. The input was an exposure-corrected Chandra ACIS image of M87 in the 3.0-7.0 keV band, where the central AGN and jet were removed and interpolated over using the surface brightness in nearby pixels (see \citealt{XRISM_M87I} for details -- hereafter \citetalias{XRISM_M87I}). This image, as well as the corresponding C, E, SW, and NW sky regions, are shown in the left panel of Fig. \ref{fig:chandramap}. 
To model nonthermal emission from the AGN, we computed an \texttt{arf} using the POINT SOURCE mode, specifying the coordinates of the center of M87 as input. To model emission from low mass X-ray binaries (LMXBs) intrinsic to each pointing, we further compute \texttt{arfs} using the POINT SOURCE mode, specifying as input the coordinates of the aim point of each observation. Note that the distribution of resolved and unresolved LMXBs is, in principle, spatially extended in a complex manner, such that finding a correct input for this raytracing is non-trivial. This introduces uncertainties in the scattering, hence the best fit LMXB fluxes should be interpreted with caution throughout this manuscript.  

Lastly, the non-X-ray background (NXB) was extracted from a database of XRISM/Resolve night-Earth data using the task \texttt{rslnxbgen}. These event files were filtered with the same criteria as the observations. Because the core of Virgo is typically much brighter than the NXB (even in the offset pointings), and the current NXB exposure time is still limited, we chose not to impose any limits on the cut-off rigidity. For all pointings except SW2, we used version 1 of the NXB database; SW2 was observed half a year later than other ObsIDs, therefore we used version 2 of the NXB database in this case.

For each pointing, we further calculated the heliocentric velocity corrections using the \texttt{barycen} HeaSOFT tool, which uses information from the orbit auxiliary files. This is particularly important for the two SW observations, which were taken 6 months apart, hence have a relative velocity shift of $\sim$55~km/s mainly due to the orbital motion of the Earth at the two different epochs. The velocity correction applied (added to) each pointing is summarized in Table \ref{tab:obs}.

\section{Spectral modeling}\label{sect:specmo}

We present results obtained using the spectral fitting package \texttt{XSpec} (version 12.15.0, \citealt{Xspec}). 
The ICM is modeled using one or several thermal components in collisional ionization equilibrium (CIE), using the atomic line emission models from AtomDB version 3.1.2 or SPEXACT version 3.08.01. All spectra were optimally binned \citep{KaastraBleeker} before fitting, and the C-statistic was used for minimization \citep{cash1979}. All abundances are reported with respect to the reference proto-solar abundance table of \cite{lodders2009}. 
The central AGN and jet were modeled as a power-law. Unless noted otherwise, the parameters of this model were fixed to values obtained from Chandra, namely the photon index is $\Gamma=2.4\pm0.1$, and the unabsorbed flux in the energy range $2-10\,$keV is $1.52^{+0.10}_{-0.09}\times10^{-12}\,\textrm{erg}\,\textrm{cm}^{-2}\,\textrm{s}^{-1}$ (see \citetalias{XRISM_M87I} for details). 
Unless specified otherwise, the LMXB contribution is modelled as an additional power-law, with index fixed at 1.7, and free normalization.
All the above components are absorbed by a Galactic hydrogen column density fixed at $2\times10^{20}$ cm$^{-2}$ (computed as the weighted average including both neutral and molecular hydrogen, using the method of \citealt{willingale2013}). 
The NXB spectra for each pointing were fit independently in the 1.7–17 keV energy range, using a diagonal \texttt{rmf} and a model consisting of a power-law component along with Gaussian profiles for the known detector emission lines. The NXB model was then fixed and included in all spectral fits shown in Sections \ref{sect:bands} and \ref{sect:multiT_c}-\ref{sect:ESW}. 

\section{Results}

\subsection{Velocity structure in different energy bands}\label{sect:bands}

We start by fitting the data obtained from the full field of view of all four regions mapped by XRISM/Resolve. We include the Galactic absorption and NXB model as described above, and fit a single temperature \texttt{bvapec} model. In the central pointing, we further added a power-law model to describe the central AGN and jet, but no LMXB component was included in these fits\footnote{we have verified that adding an LMXB component whose flux is fixed at the best-fit value reported in Table \ref{tab:2T2v_joint} does not impact the velocities reported for the offset pointings in the `hard' band, which has the highest contrast between non-thermal and thermal continuum.}.

We fit independently three different energy bands: 
\begin{itemize}
    \item the `soft' band, 2--3~keV, containing mainly emission from Si and S lines;
    \item the `medium' band, 3--4.2 keV, containing mainly Ar and Ca lines;
    \item and the `hard', 6--7 keV band, containing only emission from Fe He-$\alpha$.
\end{itemize}
In each case, the gas temperature, redshift, velocity dispersion, emission measure, and abundances of elements whose lines dominate in the band (as listed above) were left as free parameters. Abundances of elements previously fitted in a softer band were frozen to that value (e.g. in the 3--4.2 keV band, Si and S abundances were fixed to the values from the 2--3 keV fit), while all other abundances were fixed to 1.0 Solar. We used for each pointing the extended \texttt{arf} describing emission from that sky region which is recorded in the corresponding detector region (e.g. from the SW sky sector to the SW field of view), but for this exercise no spatial spectral mixing (e.g. emission originating in the central sky region, detected in the SW) is accounted for. This allows a first exploration of the data, free from potential systematic uncertainties that may be associated with raytracing and source modeling. 

\begin{table}[]
    \caption{Velocity dispersion $\sigma_v$, and heliocentric-corrected line of sight velocity $v_{\rm l.o.s.}$, for each of the investigated regions and energy bands. } 
    \centering
    \begin{tabular}{cccc}
    Region & Energy & $v_{\rm l.o.s.}$ & $\sigma_v$ \\
     & band & (km/s) & (km/s) \\
    \hline
      C   & 2--3 keV & -41±10 & 181±15 \\
      C   & 3--4.2 keV & -34±11  & 169±16 \\
      C   & 6--7 keV & 13±12 & 153±16 \\
   \hline   
      E   & 2--3 keV & -6±16 & 147(-29,+26) \\
      E   & 3--4.2 keV & 0±14 & 91(-35,+28) \\
      E   & 6--7 keV & -26±11 & 81(-20,+22) \\
    \hline
      SW   & 2--3 keV & -104±16 & 156(-32,+29) \\
      SW   & 3--4.2 keV & -87±16  & 108(-30,+27) \\
      SW   & 6--7 keV & -65±11 & 91±17 \\
    \hline
      NW  & 2--3 keV & -73±15 & <84 (2$\sigma$ u.l.)\\
      NW   & 3--4.2 keV & -41±14 & 77(-44,+30) \\
      NW   & 6--7 keV & -31±9 & 58(-20,+17) \\
    \end{tabular}
    \label{tab:bands_vel}
\end{table}

\begin{figure*}
    \centering
    \includegraphics[width=0.9\linewidth]{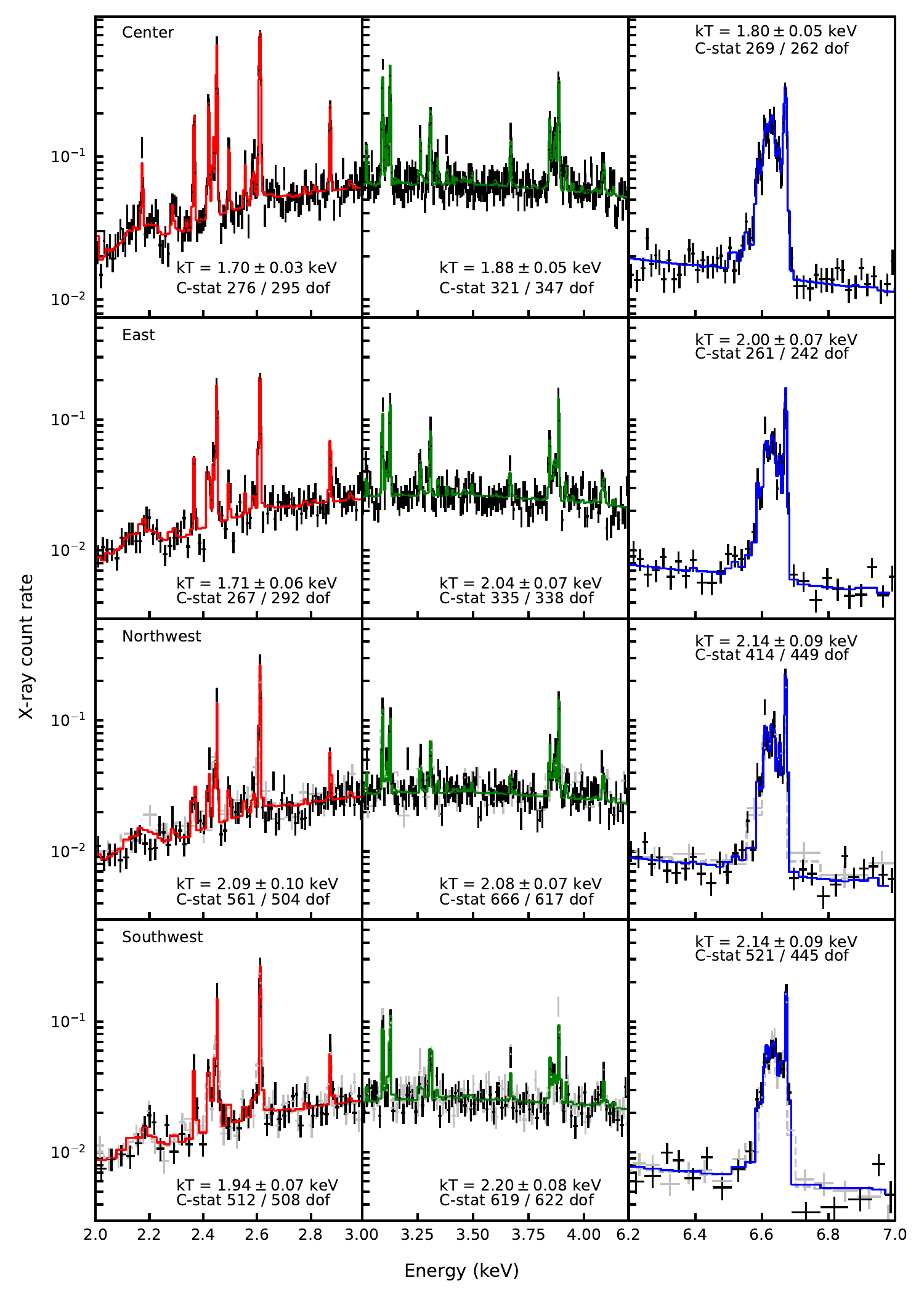}
    \caption{Spectra and best-fit single temperature models in the three energy bands considered here. When two ObsIDs cover the same region, the one with shorter exposure time is shown in silver.}
    \label{fig:spec}
\end{figure*}

The two NW pointings were fit in parallel, with all ICM parameters assumed to be identical between the two. We apply an average heliocentric correction of -25.6 km/s to the resulting line of sight velocity. The two SW pointings were fit in parallel, with all ICM parameters identical to each other, except for the redshift of SW2 which was tied to be blueshifted by 55~km/s compared to that of SW1, reflecting the different heliocentric corrections between the two. The best fit line of sight velocity was then adjusted using the heliocentric correction of SW1.

The spectra and best fit models for each of the energy bands and each of the pointings considered here are shown in Figure \ref{fig:spec}. The best-fit temperature and fit statistic are also shown as labels in each plot. The fits are generally acceptable, with all C-stat/d.o.f.$<1.2$. We notice, however, that the temperatures measured in the soft band are systematically lower than those measured in the medium and hard bands, for all pointings except the NW. This is expected, since the C, E, and SW pointings are known to contain multi-phase gas, noticeable even in low spectral resolution CCD data \citep{molendi2002,simionescu2008}. A more detailed investigation of the multi-temperature structure of the ICM based on the observed ratios of strong spectral lines, as well as the metal abundance ratios, is the subject of separate publications (Martin et al., in prep; XRISM Collaboration, in prep). 

\begin{figure*}
    \centering
    \includegraphics[width=\linewidth]{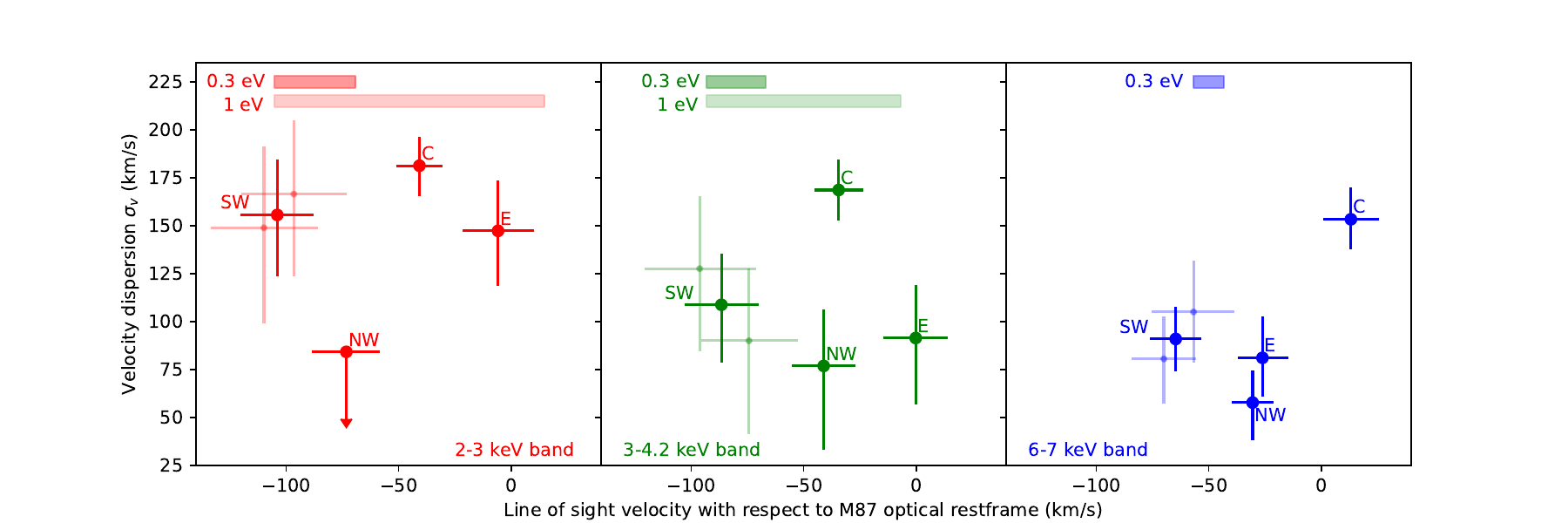}
    \caption{Velocity dispersion and line of sight velocity for each of the four pointings, for various fitted energy bands. Left: 2.-3. keV, middle: 3.0-4.2 keV, right: 6.-7. keV. For the SW pointing we also show separately, in a lighter shade, the results for SW1 and SW2. The darker bars at the top represent gain uncertainties of 0.3 eV, converted to velocity units at 2.5, 3.5, and 6.7 keV. The lighter bars in the left and middle panels additionally illustrate a gain error of 1.0 eV.}
    \label{fig:velocity_bands}
\end{figure*}

Here, we focus instead on the velocity structure of the ICM. In Figure \ref{fig:velocity_bands} and Table \ref{tab:bands_vel}, we show the velocity dispersion, and heliocentric-corrected line of sight velocity, for all four pointings and in each of the three energy bands. Note that only statistical errors are quoted here; we caution the reader that systematic uncertainties can play an important role, as discussed in detail in Section \ref{sect_gain}.

Several noteworthy observations can be made based on these results:
\begin{itemize}
    \item in the hard band, predominantly tracing the dynamics of the hotter atmosphere of M87, the central pointing stands out as having a higher velocity dispersion than all three offset observations. This is consistent with the strong peak in velocity dispersion in the core of M87 already reported in \citetalias{XRISM_M87I}.
    \item in the hard band, the velocity dispersions in pointings corresponding to the X-ray bright arms are not significantly higher than the NW direction. The extended E and SW AGN radio lobes therefore do not appear to have a measurable impact on the dynamics of the hotter ICM at their respective azimuths (see Section \ref{sect:disc_hot} for further discussion).  
    \item also in the hard band, all three offset pointings are blueshifted with respect to the central one. 
    The most blueshifted pointing is the SW arm, whose velocity is marginally higher (in absolute value) compared to both the E and NW by $|\delta v_{\rm l.o.s.}|$ of 39±16 and 34±14 km/s, respectively. 
    \item this picture starts to change, as we move towards lower energies. In the medium band, the velocity dispersion of the central pointing remains higher than all three offsets, but the redshifts of E and SW begin to separate. The SW arm is blueshifted and the E is redshifted compared to the average of C and NW, which are consistent with each other. 
    \item this trend is further enhanced in the soft band, which should more closely trace the dynamics of the cooler gas in the E and SW arms. In this energy range, the velocity dispersions of the E and SW pointings increase, becoming comparable with the center, and their redshift difference is even larger than the medium band. Meanwhile, in the relaxed NW offset, only an upper limit on the velocity dispersion can be computed, which is lower than that in the center and the arms.
\end{itemize}

Overall, this suggests that the dynamics of the cooler, uplifted gas known to be present in the E and SW pointings, may differ from that of the hotter ambient ICM.  

\subsection{Systematic uncertainties related to the calibration of the Resolve energy scale and line-spread function}\label{sect_gain}

In the 5.4–9~keV range, the XRISM/Resolve team recommends a base systematic energy-scale uncertainty of ±0.3~eV \citep{Eckart2025}, derived from semiannual checks of on-board calibration sources. This is driven by the Mn K$\beta$ line at 6.5~keV appearing at higher energies for both flat-field and point-source pixel weighting. An additional term must be added to account for time-dependent corrections per observation. For the six M87 observations, the largest residual offset in the energy-scale calibration reports was –0.232 eV (OBSID 300015010), giving an overall uncertainty of ±0.38 eV near 6 keV. At 6.7 keV, such uncertainty corresponds to ±17~km/s.

Below 5.4 keV, detailed pixel-level calibration has not been possible due to reduced soft-band effective area with the gate valve closed. The XRISM/Resolve team advises a conservative ±1 eV uncertainty, based on a +1 eV offset of the Si instrumental line in stacked data. Similar positive offsets occur in Al and Si lines from the NXB database and atmospheric Ar line during solar flares, but these fluorescence lines are unsuitable for absolute calibration better than 1 eV. Their energies vary by tenths of eV in the literature, and line shapes are affected by multiple unresolved satellite lines, the individual strengths of which depend on the chemical environment of the emitting atom and the process responsible for creating the precursor hole. Additionally, the weakness of these incidental lines precludes fitting individual pixels and dividing the long integration times by housekeeping parameters such as the temperature of the read-out electronics. 
All of this leaves us with the unsatisfying conclusion that it is not currently possible to quantify the energy-scale errors below 5.4 keV to better than the qualitative 1~eV recommendation, and that, in the 1.5 -- 3 keV regime, the data suggest that errors may be biased in the positive (blue-shifted) direction. At 2.5 keV, ±1~eV corresponds to ±120 km/s, which is larger than most velocity offsets reported in Table \ref{tab:bands_vel}. 

Regarding line broadening uncertainties, the energy-dependent Gaussian kernel of the line-spread function (LSF) of each pixel was updated based on measurements in orbit \citep{Leutenegger2025}. The LSF is characterized by the quadrature sum of a baseline term and a term that scales linearly with energy. Both terms were updated based on processed pulse-free records and the resolution at 5.9~keV, but it has not been possible to verify the widths at other energies. 
Statistical uncertainties on the coefficients imply negligible impact on the velocity dispersion: ±0.15 eV at 6.7 keV adds ±6 km\,s$^{-1}$ to $\sigma_v$=40 km\,s$^{-1}$ and ±2 km\,s$^{-1}$ to 140 km\,s$^{-1}$; ±0.06 eV at 2.5 keV adds ±4 km\,s$^{-1}$ for $\sigma_v$ =140 km\,s$^{-1}$.  
However, misalignment of energy scales across pixels or time could propagate to LSF errors. The energy scales are pinned at 5.9 keV, thus the 6-7~keV range should be well aligned. Preliminary tests analyzing gain-tracking data in sets of 100~ks show a low Mn K$\beta$ dispersion of $\sim$0.2 eV across pixels and $\sim$0.02 eV across integrations. The larger term broadens lines by 9 km\,s$^{-1}$ at 6.5 keV, which results in a difference of only 1 km\,s$^{-1}$ when added in quadrature to a turbulence of 40 km\,s$^{-1}$. Nonetheless, once again, the effect could be more severe at energies that are farther away from the 5.9 keV anchor point. If the conservative ±1~eV energy-scale errors at low energies induce a corresponding 1~eV broadening due to scale misalignment, then a measured dispersion of 120 km/s at 2.5~keV could be consistent with 0.

We therefore cannot exclude the possibility that energy-scale systematic uncertainties of up to ±1~eV are responsible for (at least part of) the differences in redshift and broadening observed between the soft, medium, and hard bands shown in Figure \ref{fig:velocity_bands}. To investigate whether the systematic errors in the M87 velocities may be lower than this, we are left with self-checks within the present dataset itself. This gives us several reasons to believe that the impact of the detector gain calibration is unlikely to dominate the results: 

\begin{itemize}
    \item as mentioned, the SW arm was observed twice, at an interval of 6 months. Notably, the two pointings, while covering the exact same sky area, have a roll angle that is different by exactly 180 degrees, such that different pixels cover different parts of the sky. In Figure \ref{fig:velocity_bands}, we have also added the measurements obtained separately from SW1 and SW2 (with the respective appropriate heliocentric corrections), which are perfectly consistent with each other, with a 2--3 keV l.o.s. velocity difference of 13±33 km/s. This argues that the impact of the time dependence or position dependence of the gain calibration is low.
    \item we have further split the C and E observations into two (roughly equal) time intervals, to check for any short term gain variations. The best fit 2--3 keV redshifts of the two halves of the C observation differ by only 29±20 km/s, and for the E by 38±33 km/s, hence not revealing any significant variations. The velocity dispersions are also consistent (within $1\sigma$ statistical error) between each pair of sub-exposures. Note that, in all cases (including SW1 and SW2), the temperature and abundance(s) of each half-pointing were coupled to each other during the fit, while the norm, redshift, and velocity dispersions were allowed to vary independently.  
    \item based on prior knowledge from CCD data, the NW offset should be remarkably close to isothermal (see discussion in \citealt{million2010}). It is noteworthy that exactly this pointing is well behaved also in terms of its velocities, exhibiting the most stable $v_{\rm l.o.s.}$ as a function of energy band, and showing no increase in $\sigma_v$ towards softer energies, unlike the E and SW arm offsets. It would be a surprising coincidence if this were due to random gain variations only. The NW spectrum does exhibit a slight trend in $v_{\rm l.o.s.}$, which is blueshifted by $|\delta v_{\rm l.o.s.}|$ of 42±17 km/s in the soft band (but only 10±17 km/s in the medium band) compared to the hard band. We do therefore see potential evidence of systematic energy-scale errors. Although the offset is less than 1 eV, the direction is consistent with the observed positive offsets of the Al and Si instrumental lines and Ar atmospheric line, keeping in mind all the caveats mentioned in the beginning of this section.   
    \item finally, and perhaps most speculatively, the sign of the velocity difference between E and SW in the soft band is worth noting. 
    Studies of polarized radio emission from the southern lobe have shown that the Faraday depth towards it is likely to be small \citep{andernach1979}. On smaller, kpc scales, the well known jet of M87 is also pointed westward, showing apparent superluminal motion indicating an angle of $<19^\circ$ from the line of sight \citep{biretta1999}. \cite{werner2010} have used these facts to argue that the SW X-ray arm should be oriented towards us, while the E arm is pointed away from us. In this case, it would be exactly expected for the softer X-ray emission in the SW to be blueshifted compared to the E, as observed by XRISM. Once more, it would be a remarkable -- though of course not impossible -- coincidence for this signal to be purely due to the gain calibration of the detectors.
\end{itemize}    

Another way to investigate the presence of velocity shifts between the hotter and cooler gas phases would be to consider separately Helium-like and Hydrogen-like transitions, especially for lighter elements like Si and S. These are relatively close in energy, reducing the impact of gain calibration uncertainties; cooler gas is expected to have brighter Helium-like transitions, while Hydrogen-like lines trace better the hotter gas. Unfortunately, the present data quality, especially given the closed gate valve, does not allow us to detect statistically significant differences between the widths or shifts of Si Ly-$\alpha$, Si He-$\beta$, S Ly-$\alpha$, and Si He-$\alpha$ (the four strongest lines in the soft band).

In summary, systematic energy-scale uncertainties are estimated at $\pm$0.3--0.4 eV in the 5.4--9 keV range and up to $\pm$1 eV below 5.4 keV, where calibration is less constrained. Around 2.5~keV, this corresponds to velocity errors as large as $\pm$120 km\,s$^{-1}$ on the redshift, which can also propagate into an increased line broadening. Internal consistency checks within the M87 dataset itself suggest that gain calibration effects are unlikely to drive the observed velocity differences between different energy bands, although systematic uncertainties at a lower level of tens of km\,s$^{-1}$ are likely present and difficult to quantify at this time. 

\subsection{Multi-temperature, multi-velocity fits}

Given the indications for multi-velocity structure in the regions of M87 containing multi-phase gas, as argued above, we proceed to fit the spectra in a broad, 1.7--9.0 keV band, including multiple thermal models. The choice of energy band is motivated at the low end by including the Si He-$\alpha$ line, which although having low statistics due to the attenuation by the gate valve, should be a sensitive probe of the cooler X-ray gas phases. At the high end of the band, the NXB becomes comparable to the ICM flux in the offset pointings above 9.0 keV.

\subsubsection{The central region}\label{sect:multiT_c}

We fit the spectrum of the central pointing with a two temperature model, plus two additional powerlaw components describing the contribution from the central AGN and jet, as well as LMXBs (see Section \ref{sect:specmo}).
The temperatures, normalizations, redshifts, and velocity broadening of each ICM component were treated as free parameters in the fit. 
Throughout the text, the normalizations of the thermal components are reported with respect to a convenient multiple of the standard Xspec units, which are defined as 
\begin{equation}\label{eqn:xspecnorm}
    Norm = \frac{10^{-14}}{4\pi[D_A(1+z)]^2}\int n_e n_H dV,
\end{equation}
where $D_A$ is the angular diameter distance to the source (cm), $dV$ is the volume element (cm$^3$), and $n_e$ and $n_H$ are the electron and hydrogen number densities (cm$^{-3}$), respectively. 

The abundances of Si, S, Ar, Ca, Cr, Fe, and Ni were coupled between the cooler and hotter phases, and also allowed to vary. Due to the centrally peaked X-ray surface brightness profile of M87, the scattering of photons from outside the central field of view (FoV) into this region is negligible (below 5\% of the intrinsic flux) and we therefore do not include this in our model.

The best fit results are summarized in Table \ref{tab:2T2v_fovfits}, except for the metal abundances which are discussed elsewhere (XRISM Collaboration, in prep). Notably, we find that the cooler gas phase with a temperature of around 1.3~keV is blueshifted with respect to the hotter 2.0~keV component by -178(-51,+72)~km/s, i.e. at a $\sim$2.5 $\sigma$ statistical significance level. However, investigating the impact of potential instrumental calibration and model systematics (see Appendix \ref{app:center}), we conclude that this velocity shift between the hotter and cooler gas is very sensitive to assumptions about the gain calibration at lower energies. We therefore choose not to further discuss its physical interpretation (but do include the different best fit velocities in order to model light scattered from the center into the offset pointings in the next sections).
The velocity of the hotter, dominant component (with an emission measure ratio of 2:1 with respect to the cooler one) is roughly consistent with the optical restframe of M87. It is also consistent with the velocities reported in \citetalias{XRISM_M87I}, which specifically focused on the dynamics of the hotter ICM by limiting the analysis to an energy band above 3 keV. The velocity dispersions of both thermal phases are in excellent agreement, at around 150 km/s. 

\begin{table*}[]
    \caption{Results of fitting the full FoV spectra of individual pointings expected to show multi-phase substructure.}
    \label{tab:2T2v_fovfits}
    \centering
    \begin{tabular}{c|c|c|c}
       Region  & C & E & SW  \\
       \hline
       kT$_1$ (keV) & 1.30(-0.07,+0.17) & 1.50(-0.09,0.11) & 1.77 (-0.39,0.08) \\
       norm$_1$ & 1.54(-0.34,+0.99) & 2.49(-0.46,+0.54) & 2.33(-1.58,+0.36) \\
       v$_1$ (km/s)  & -160(-42,+71) & 50(-27,+15) & -152 (-198,+27) \\
       $\sigma_{v1}$ (km/s) & 153(-47,+40) & 172(-36,+34) & $<215$ (2$\sigma$ u.l.) \\    
        \hline
       kT$_2$ (keV) & 1.97(-0.07,+0.17) & 2.44±0.14 & 2.72(-0.47,+0.16) \\
       norm$_2$ & 2.91(-1.01,+0.31) & 1.84(-0.55,+0.42) & 1.55(-0.35,+1.29) \\       
       v$_2$ (km/s) & 18(-12,+29) & -38(-12,+11) & -45(-15,+15) \\
       $\sigma_{v2}$ (km/s) & 151(-25,+14) & $<67$ (2$\sigma$ u.l.) & $<97$ (2$\sigma$ u.l.) \\ 
        \hline
       LMXB flux & -11.81(+0.06,-0.05) & -12.44(-0.17,0.11) & $<-12.56$ (2$\sigma$ u.l.) \\
         \hline
       C-stat/d.o.f. & 1922/1970 & 1840/1866 & 3685/3410 \\
    \end{tabular}
    \tablefoot{All velocities are reported with respect to a rest frame heliocentric redshift of 0.00428. The LMXB flux is given in log(erg/s/cm$^2$) in the 2--7 keV band. The normalizations of the thermal components are given in 1e-2 Xspec units (Eqn. \ref{eqn:xspecnorm}).}
\end{table*}

\subsubsection{The East and Southwest arms}\label{sect:ESW}

We proceed to fit the full FoV of the E and SW observations, assuming that the local ICM is described by a two temperature model. We initially make the same assumptions as for the central pointing, namely we leave free the temperatures, normalizations, redshifts, and velocity broadening of each thermal component, and couple their Si, S, Ar, Ca, Cr, Fe, and Ni abundances. Here, we further include scattering from the central ICM and AGN. We compute an \texttt{arf} in IMAGE mode (\texttt{imgarf}), performing the raytracing from the central sky region to the E and SW detector regions, respectively, and fold the best fit 2T model from the previous section with these effective area files. We further compute an \texttt{arf} in POINT SOURCE mode (\texttt{psarf}), which contains the raytracing from the central AGN and jet to the E and SW detector regions, and fold the AGN power-law with this response. Lastly, we include a local LMXB component as a separate power-law model folded with a POINT SOURCE arf centered at the aim point of each observation. In detail the model is:
\texttt{phabs*(bvvapec$_{P1}$+bvvapec$_{P2}$)*imgarf(skyP$\rightarrow$detP) + phabs*(bvvapec$_{C1}$+bvvapec$_{C2}$)*imgarf(skyC$\rightarrow$detP) + phabs*powerlaw$_{\rm AGN}$*psarf(AGN$\rightarrow$detP) + phabs*powerlaw$_{\rm LMXB}$*psarf(aimP$\rightarrow$detP) + NXB}, with P in [E,SW]. The parameters of the AGN and the central model for the scattered light are all frozen. Note that we do not include scattered LMXB flux from the center to the E and SW pointings; however, since the local LMXB normalization is a free parameter, it can adjust to include this contribution. The LMXB fluxes in SW1 and SW2 were tied to each other.

The best fit results are summarised in Table \ref{tab:2T2v_fovfits}. Based on these, we can make the following observations:
\begin{itemize}
    \item the fit cannot simultaneously constrain the velocity dispersions of each of the two thermal phases. However, in the E pointing, there are hints that the velocity dispersion of the cooler gas is inconsistent with that of the hotter phase. Namely, $\sigma_{v1}$ is higher than the upper limit of $\sigma_{v2}$, with a statistical confidence of 2.9$\sigma$. Interestingly, the velocity dispersion of the cooler gas in the E pointing is consistent with the velocity dispersions of both thermal phases in the central pointing. 
    In the SW region, neither velocity dispersion is constrained when both are left free. For the hotter phase, the upper limits on $\sigma_{v2}$ are consistent in both arm regions with the value measured in the relaxed NW offset (`region 3' of \citetalias{XRISM_M87I}.) This is in line with the results based on the 6--7 keV band fits shown in Fig. \ref{fig:velocity_bands}.
    \item the line of sight velocities of the hotter thermal component, $v_2$, are perfectly consistent between the E and SW, and also consistent with the bulk velocity of the NW region (-33±11 km/s; \citetalias{XRISM_M87I}). Meanwhile, the cooler gas in the E is redshifted by 88(-29,+19) km/s, and that in the SW is blueshifted by 107(-31,+199) km/s, with respect to the hotter phase in the corresponding region. This means that the cooler gas in the E is redshifted by 202(-38,+199) km/s with respect to the cooler gas in the SW. Again, this is consistent with the hints obtained from Fig. \ref{fig:velocity_bands}. 
    The fit uncertainties are large and asymmetric, particularly in the SW, likely because the limited statistics at soft energies makes it difficult to separate the two thermal components.  
    \item the temperatures of the cooler component are statistically consistent between the E and SW pointings. The temperatures of the hotter component in the on-arm pointings also agree with each other, and are consistent within about 1$\sigma$ with the single-temperature model measurement of the relaxed NW offset (2.26$^{+0.09}_{-0.08}$ keV reported in \citetalias{XRISM_M87I}).
\end{itemize}

\subsubsection{Joint fitting of the offset pointings}\label{sect:joint}

The multi-temperature, multi-velocity fits presented above support a picture where the cooler, uplifted gas phase in the E and SW regions has a higher velocity dispersion and different line of sight velocities compared to the hotter phase, as already hinted by the results obtained from different energy bands in Section \ref{sect:bands}. However, many fit parameters are poorly constrained and/or have asymmetric error bars, indicating degeneracies which cannot be broken with the present data quality, if individual pointings are modeled independently. This problem can be alleviated by performing a joint fit of multiple XRISM observations, taking advantage of the statistical consistency of many of the parameters reported in the previous subsection. To this end, we fit all the offset observations (E, SW, and NW) in parallel with the following assumptions (hereafter `model A'):
\begin{itemize}
    \item the temperature and velocity broadening of the cooler component are coupled between the E and SW pointings. 
    \item the NW pointing has no cooler gas, hence the normalization of this component is set to 0.
    \item the temperature and velocity broadening of the hotter component are coupled between the E, SW, and NW pointings. 
    \item the abundances of Si, S, Ar, Ca, Cr, Fe, Ni are free to vary in each pointing independently; where two thermal components are present (in E and SW), the respective abundances between the two phases are coupled (Si$_{\rm 1,E}$=Si$_{\rm 2,E}$, S$_{\rm 1,E}$=S$_{\rm 2,E}$,...,Si$_{\rm 1,SW}$=Si$_{\rm 2,SW}$, etc.)
    \item the redshifts and normalizations of all five thermal components (cooler gas in the E and SW, hotter phase in E, SW, NW) are treated as free parameters.
    \item the LMXB fluxes are treated as free parameters for each pointing individually; the LMXB flux in SW1 and SW2 are assumed to be equal.
\end{itemize}
For this fit, in order to keep the already large number of parameters and response files to a manageable level, we chose to directly subtract the NXB rather than model it as in the previous sections. 

The results of this fit are shown in Table \ref{tab:2T2v_joint}, and plots comparing the model with the observed spectra are shown in Appendix \ref{app:spec}. The fit confirms yet again the presence of a line of sight velocity difference between the various gas phases. The cooler gas in the E is redshifted by 118(-34,+42) km/s compared to the hotter phase in the same pointing, while in the SW the cooler gas is blueshifted at -262(-77,+58)~km/s with respect to its surrounding hot ambient. 
Moreover, the velocity dispersion of the cooler gas is $4.4\sigma$ higher than the velocity dispersion of the hotter gas (although this relies on the uncertain assumption that $\sigma_{\rm v,cool}$ is the same in both arm regions).
We note that, while all of the velocity parameters (shift and broadening) remain statistically consistent with the fits performed for the individual pointings shown in Table \ref{tab:2T2v_fovfits}, the uncertainties have decreased significantly, and error intervals are considerably more symmetric, thanks to the improved constraining power of the joint fitting. 

\begin{table}[]
    \caption{Results of jointly fitting all three off-center observations of M87, as described in Section \ref{sect:joint}.} 
    \label{tab:2T2v_joint}
    \centering
    \begin{tabular}{c|c|c|c}
       Region  & E & SW & NW \\
       \hline
       kT$_1$ (keV) & \multicolumn{2}{|c|}{1.50(-0.09,+0.11)} & - \\
       \hline
       norm$_1$ & 1.97$_{-0.34}^{+0.34}$ & 0.90$_{-0.24}^{+0.29}$ & 0 (fixed) \\
       v$_1$ (km/s) & 76(-32,+40) & -322(-74,+57) & - \\
       \hline
       $\sigma_{v1}$ (km/s) & \multicolumn{2}{|c|}{153(-21,+40)} & - \\    
        \hline
       kT$_2$ (keV) & \multicolumn{3}{|c}{2.23±0.04}  \\
       \hline
       norm$_2$ & 2.32$_{-0.35}^{+0.31}$ & 3.03$_{-0.25}^{+0.22}$  & 3.83$_{-0.09}^{+0.09}$\\       
       v$_2$ (km/s) & -42±12 & -60±10 & -46±7 \\
       \hline
       $\sigma_{v2}$ (km/s) & \multicolumn{3}{|c}{43(-11,+14)}  \\ 
        \hline
       flux$_{\rm LMXB}$ & -12.39$_{-0.12}^{+0.09}$ & -12.62$_{-0.22}^{+0.15}$ & -12.50$_{-0.18}^{+0.13}$ \\
         \hline
       C-stat/d.o.f. & \multicolumn{3}{|c}{9172/8710}  \\
    \end{tabular}
    \tablefoot{All velocities are reported with respect to a rest frame heliocentric redshift of 0.00428. The LMXB flux is given in log(erg/s/cm$^2$) in the 2--7 keV band. The normalizations of the thermal components are given in 1e-2 Xspec units (Eqn. \ref{eqn:xspecnorm}).}
\end{table}

\subsubsection{Model uncertainties on the offset pointing joint fits}\label{sect:joint_mosys}

\begin{figure*}
    \centering
    \includegraphics[width=\linewidth]{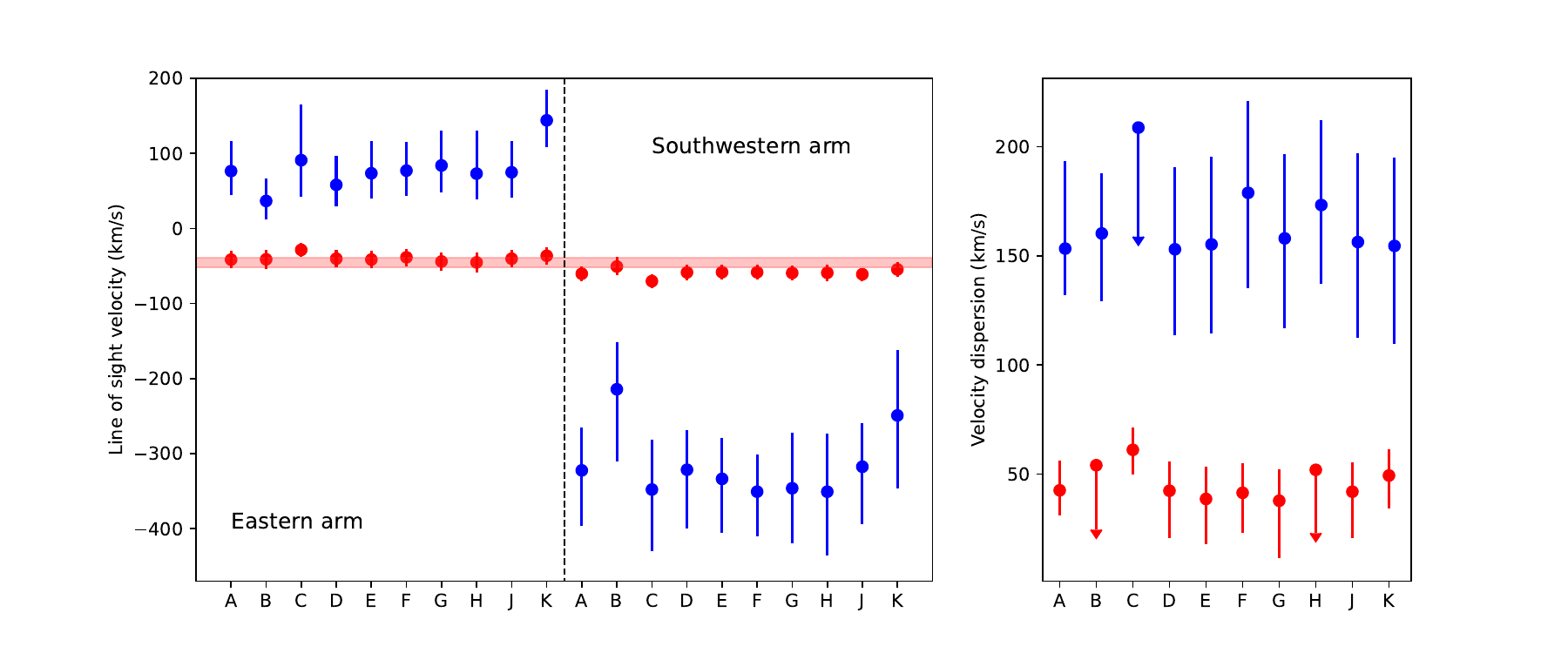}
    \caption{Left: line of sight velocity of the cooler (blue) and hotter (red) gas phases, for each of the two arm pointings and each of the various models and assumptions discussed in Sections \ref{sect:joint}-\ref{sect:joint_mosys}. The 
    $1\sigma$ statistical confidence interval of the l.o.s. velocity of the NW offset is shown as a light red band for comparison. This parameter is insensitive to the model variations, with a standard deviation of only 0.5 km/s between the different models presented.
    Right: corresponding velocity dispersions of the cooler and hotter gas.}
    \label{fig:joint_fits}
\end{figure*}

In order to verify whether our results are sensitive to the assumptions listed above, in addition to our fiducial `model A', we further fit the data with a range of other spectral models, modifying or relaxing some of those assumptions as follows:

\begin{itemize}
    \item Model B: as model A, but the temperatures of the hotter phase are allowed to vary independently in E, SW, and NW.
    \item Model C: as model B, but the temperatures of the cooler phase in E and SW are frozen to their respective best-fits obtained from XMM-Newton EPIC data. 
\end{itemize}
To this end, we re-analysed XMM ObsIDs 0803670501 and 0803670601 using SAS v22 and the pipeline described in \cite{rossetti2024}. The observations total around 185~ks and 140~ks of net exposure for the MOS1/2 and pn detectors, respectively. We extracted spectra from regions closely matched to the XRISM E and SW fields of view. The spectra were fit with a Galactically absorbed two temperature plus LMXB powerlaw model (plus additional components for the sky and detector backgrounds), with abundances of O,Ne,Mg,Si,S,Ar,Ca,Fe,Ni coupled between the thermal components and left free. As already reported in previous studies of M87 \citep{belsole2001,matsushita2002,molendi2002}, the EPIC data suggests cooler gas phases with temperatures lower than the XRISM best fit values, namely $kT=1.130\pm0.008$~keV in the E, and $kT=1.116(-0.015,+0.013)$~keV in the SW. This is undoubtedly driven by the Fe-L line complex in the EPIC soft band, which is inaccessible to XRISM/Resolve due to the closed gate valve.

\begin{itemize}
    \item Model D: as model A, but the Si, S, Ar, Ca, Cr, Fe, Ni abundances of the cooler phase in each of the arms are fixed to the respective best fit values obtained from the Central pointing. This is justified, assuming that this represents uplifted gas.
    \item Model E: as model A, but the abundances of Si, S, Ar, Ca, Cr, Fe, Ni for the cooler phase in each of the arms are all coupled to each other, and left as one free parameter in the fit. De facto, this assumes a Solar composition of the cooler gas, with its absolute abundance allowed to vary. We obtain only loose constraints on this metallicity parameter, at 1.15(-0.33,+0.67) Solar.
    \item model F: as model A, but the velocity dispersion of the cooler gas is no longer assumed to be the same in both the E and SW arms. Instead, $\sigma_{\rm v,cool}$ in the SW arm is assumed to be the same as that of the hotter gas, while $\sigma_{\rm v,cool}$ in the E arm is left to vary completely independently. This is justified by the smooth and regular morphology of the SW filaments seen with Chandra, suggesting their turbulence may not be as high as the far less ordered ones in the E.
    \item model G: as model A, but the hotter phase is modeled using a Gaussian emission measure distribution with free width $\sigma_{\rm T,hot}$, instead of a single temperature model. Both $\sigma_{\rm T,hot}$ and the average temperature $T_{\rm mean,hot}$ are coupled between E, SW, and NW.
    \item model H: as model G, but the cooler phase is also modeled using a Gaussian emission measure distribution with free width $\sigma_{\rm T,cool}$, instead of a single temperature model. Both $\sigma_{\rm T,cool}$ and the average temperature $T_{\rm mean,cool}$ are coupled between E and SW.
    \item model J: as model A, but using SPEXACT v3.08.01 instead of AtomDB v3.1.2. 
    \item model K: as model A, but modifying all spectral response files to introduce a linear gain transformation, corresponding to no shift at 6.7 keV, and a -0.33 eV shift at 2.5 keV, prompted by the 42 km/s blueshift between the soft band and the hard band in the NW pointing.
\end{itemize}

The gas velocities obtained from models A-K are summarized and compared in Figure \ref{fig:joint_fits}. All of these models suggest significant velocity offsets between the cooler gas in the E and SW arms, while the hotter phase has a consistent line of sight velocity between all three offset pointings. This is also true when various further instrumental / calibration uncertainties are considered (NXB, level of scattering, effective area), which we elaborate on in Appendix \ref{app:offsets}. 

The smallest difference between $v_{\rm cool,E}$ and $v_{\rm cool,SW}$ is obtained for model B, but even in this case it is 3.7$\sigma$ significant with $v_{\rm cool,E}-v_{\rm cool,SW}=251(-68,+101)$ km/s. 
All models except model C also suggest that the cooler gas has a higher velocity dispersion than the hotter one. For models B and H, the velocity dispersion of the hotter gas is not constrained, and only a $2\sigma$ upper limit on its value is plotted. 

We note, however, that applying a more conservative shift of -1~eV to the \texttt{rmfs} results in the cooler gas phases in the two arms both becoming blueshifted compared to the hotter phase, and reduces their velocity difference. Our conclusions are therefore only robust if the gain calibration errors of the XRISM/Resolve detector are of similar magnitude across the entire detector band-pass.

\section{Discussion}

The XRISM/Resolve mosaic of pointings covering the core of M87 analysed in this work reveals several interesting details regarding the dynamics of the multitemperature X-ray gas, and its relation to AGN feedback. We discuss the implications for the hotter, ambient ICM in Section \ref{sect:disc_hot}, and for the cooler, uplifted gas in the X-ray arms in Section \ref{sect:disc_cool}.

\subsection{Impact of older AGN lobes on ICM dynamics}\label{sect:disc_hot}

We find that the widths and shifts of the Fe He-$\alpha$ line at 6.7 keV do not show marked differences between the three off-center pointings (Table \ref{tab:bands_vel} and Figure \ref{fig:velocity_bands}). 
Compared to the NW pointing, located away from the extended AGN lobes of M87, the SW observation is tentatively blueshifted in the hard X-ray band by only 34±14(stat)±13(sys) km/s, where the quoted systematic uncertainty corresponds to gain shifts of ±0.3 eV. Meanwhile, the redshift of the E arm pointing is perfectly consistent with the NW (within the $1\sigma$ confidence interval). The Fe He-$\alpha$ line widths in the SW and E are higher by 1.4 and 0.8 $\sigma$, respectively, compared to the NW. Hence, in the hard X-ray band, we find no significant differences between the gas dynamics in the on-lobe and off-lobe directions.

Previous work by \cite{gatuzz2022}, based on a detailed calibration of XMM-Newton EPIC/pn spectra, reported that the hot ICM within the western radio flow of M87 was redshifted, moving with a velocity of $\sim$331±200~km/s while the hot gas located within the eastern radio flow was tentatively blueshifted, with a velocity of $\sim$258±400~km/s.  
Our results do not confirm the presence of ICM bulk velocities of such a large amplitude and, if at all, would suggest a spatial gradient in the opposite direction, with the SW arm being blueshifted compared to the E. This showcases the importance of high resolution spectra in constraining ICM motions. 

The hard-band Chandra image of M87 (Figure \ref{fig:chandramap}, left) appears azimuthally symmetric, showing no significant emissivity enhancements or deficits at the positions of the E and SW `arms'. This energy band thus samples a similar gas volume and geometry across all three offset pointings, allowing the measured velocities to be directly compared without correcting for differences in effective emission scales. Consequently, the XRISM measurements based on the Fe He-$\alpha$ line at 6.7 keV -- showing no substantial variation in velocity or velocity dispersion between on-lobe and off-lobe directions -- indicate, perhaps surprisingly, that the older E and SW radio lobes in M87 do not produce a strong dynamical signature in the hotter, $\sim2$~keV ambient ICM.

A possible interpretation of this finding is that gas motions in the ICM dissipate quickly, such that dynamical signatures of the interaction with the older lobes have already been erased. 
Other scenarios, however, cannot be ruled out yet. For example, the older AGN outburst could drive motions in a more homogeneous fashion, not only along the directions of the observed radio lobes (recall that the even larger scale outer radio bubbles fill essentially the entirety of the M87 core). Alternatively, we could be observing the interaction `too soon': the cold gas uplift is still laminar and ongoing, and turbulence may still develop. Finally, it is also worth keeping in mind that the radio lobes are likely far from the plane of the sky (see also the next Subsection), and hence their kinematic signature on the ICM may become hard to detect when observed in front of or behind brighter, undisturbed ambient gas at the same projected radius. Dedicated numerical simulations are needed in order to fully explore the implications of the current velocity measurements on constraining models where radiative cooling in cluster cores is mediated by the dissipation of turbulence driven by AGN feedback.

\subsection{Geometry and energetics of cooler uplifted gas}\label{sect:disc_cool}

Before delving into this section, we reiterate that the most prominent signatures of the  cooler ($\sim$1-1.5~keV) gas are seen in the 1.7--3 keV band, which is particularly prone to systematic uncertainties related to the gain calibration. 
We consider that the difference between the velocity of the cooler gas in the E versus the SW arms is the most robust measurable quantity, since it is a differential test which relies mainly on the assumption that the gain solution did not vary between the observations; the velocity differences between the cold and hotter gas, as well as the velocity broadening of the cooler gas, are likely subject to larger uncertainties, as discussed extensively in Section \ref{sect_gain}. 

Our main conclusion regarding the dynamics of the uplifted gas is that the bright SW X-ray filaments are blueshifted with respect to the E ones, by between 250--440 km/s, depending on the exact strategy used to model the temperature structure (Figure \ref{fig:joint_fits}). This provides an important confirmation, from a gas kinematics perspective, of the existing paradigm that these substructures are the result of low entropy gas being uplifted in the wake of the AGN radio lobes which buoyantly rise through the ICM in opposite directions along the line of sight \citep[see][and references thereto]{churazov2001}. While constrained less robustly than the l.o.s. velocity differences, the velocity dispersion of the cooler gas is also consistent with this picture: an increased line width would be expected in the uplift scenario, either if the outflowing gas is more turbulent than the surrounding medium, or if we are observing a superposition of laminar motions due to many overlapping strands/filaments of colder gas. The latter interpretation is supported by the fact that the E arm, which is known to have a more complex cold gas morphology than the SW based on Chandra images, shows stronger statistical evidence of an enhanced $\sigma_{\rm v,cool}$ compared to $\sigma_{\rm v,hot}$, but a smaller velocity offset between the cooler and hotter gas. 

Radio studies have previously shown that the southern lobe has a small Faraday depth, suggesting the SW X-ray arm should be oriented towards us, while the E arm is pointed away from us \citep{andernach1979,werner2010}. In this scenario, the blueshift observed in the SW arm in relation to the E is consistent with the cooler gas being uplifted away from the M87 core (rather than falling back towards the center of the potential well). \cite{werner2010} argue, based on several considerations related to the large scale radio morphology and apparent superluminal motion of the inner jet knots, that the arms are likely oriented within 15--30 degrees from our line of sight. In this case, an observed difference of 250--440 km/s translates to a total velocity of 260--510 km/s, or roughly 35--70\% of the sound speed (which for a temperature of 2~keV amounts to 713 km/s). 

Early calculations presented in \cite{churazov2001} estimated the expected Mach number of buoyantly rising relativistic bubbles in the ICM as 0.6--0.7, with the note that this velocity is sensitive to the bubble shape, as well as the assumed compressibility of the gas, and the extent to which the bubble excites internal gravity waves that increase the drag coefficient. More detailed simulations later shown by \cite{zhang2022} indeed predict a range of bubble velocities from several tens to several hundreds of km/s, with flatter bubbles having smaller terminal velocities. The dynamics of the cooler gas measured for M87 in this work is therefore perfectly in line with theoretical expectations, under a scenario that the velocities are due to gas uplift in the wake of relativistic radio bubbles rising through the ICM at a fraction of the sound speed.

We can further estimate the age and energetics of the X-ray arms in M87 based on the results presented here; however, this requires separate values of the E and SW cool gas velocities and, again, we caution the reader that these numbers are more prone to gain systematic uncertainties than their difference. In what follows, we take "model A" as our fiducial set of kinematic parameters. For an orientation angle of 30 degrees, the deprojected length of the SW arm as seen by Chandra is 60~kpc; if the line of sight $v_{\rm cool,SW}$ is -322 km/s (Table \ref{tab:2T2v_joint}), and assuming that this arm originated in the wake of a single AGN outburst episode, and that the current velocity represents the average over the lifetime of the uplift, then the age of this outburst would be $t\sim160$ Myr -- over 10 times older than the 13~kpc AGN shock \citep{forman2017}, but comparable to the $\sim$100~Myr age estimated for the outermost radio bubbles by \cite{owen2000}.

The total mass of the cooler $\sim1$~keV gas estimated from \chandra\ mapping is $6-9 \times 10^8 M_\odot$, again depending on the inclination angle \citep{werner2010}. We calculated the mass of cooler gas in each individual arm by reanalysing the full field \chandra\ observations of M87 (Obs. IDs 352, 2707, 3717, 5826, 5827, 5828, 6186, 7210, 7211 and 7212) using the latest \textsc{ciao} software version 4.17 and following the data reduction methods described in \cite{russell2018}. 
We used the contour binning algorithm of \cite{sanders2006} to produce maps of the gas properties for a two-temperature model with free temperatures and normalizations for each phase, and one common abundance. 
This captured the dominant gas components at $\sim1$~keV and 2 keV in the regions of interest here. By selecting the $\sim1$~keV temperature component that closely traces each arm, we applied the method described by \cite{werner2010} to determine their individual cool gas masses. The SW arm gas mass was $3.2-3.5\times10^{8} M_\odot$ and the E arm gas mass was $4-4.4\times10^{8} M_\odot$, where the range reflects orientation angles between $15-30^\circ$. The l.o.s. velocities of 76$^{+40}_{-32}$ and -322$^{+57}_{-74}$ km/s in E and SW measured here with XRISM translate to total velocities of approximately 88 and -372 km/s for a 30 degree inclination angle; together with the gas masses estimated above, this implies a kinetic energy of the uplift of $\sim5\times 10^{56}$ erg. Using the updated gas masses, together with the total mass profile reported by \cite{churazov2008}, we can also estimate the gravitational potential energy needed to uplift the gas to be $\sim3.5\times 10^{57}$ erg (see also \citealt{simionescu2008} who reported a similar value of $4\times 10^{57}$ erg from XMM-Newton mapping, though using now outdated atomic data for the multi-temperature fitting).

Therefore, the kinetic energy of the uplift corresponds to $\sim$14\% of the gravitational potential energy of the same gas, i.e. gas motions encompass only a small fraction of the mechanical energy associated with feedback. Note that this conclusion necessarily relies on the approximation that all of the cooler gas in each arm has the same physical properties. We have shown in Figure \ref{fig:joint_fits} that the exact assumptions on the thermal structure and overall abundance of the uplifted gas do not significantly impact our measurements, but having a range of velocities and abundances, as opposed to a single value, could affect the measured kinetic energy. Unfortunately it is not possible to estimate this effect quantitatively with current data and existing models.

\section{Conclusions}

We present the first, coarse azimuthal mapping of gas velocities in M87 using XRISM/Resolve, focusing on the eastern and southwestern X-ray arms shaped by AGN feedback. Our main findings are:
\begin{itemize}
\item The hotter ICM phase, traced by Fe He-$\alpha$ emission, shows low velocity dispersion and no significant line-of-sight velocity differences between the arms and the relaxed NW region. This suggests that the older AGN lobes do not currently drive strong motions in the ambient hot gas.
\item The cooler, uplifted gas phase exhibits distinct kinematic signatures, with the SW arm blueshifted by 250–440 km/s relative to the E arm. This supports the scenario where AGN outbursts have uplifted low-entropy gas along opposite directions of the line of sight.
\item The cooler gas in the X-ray `arms' also tentatively shows higher velocity dispersion than the hot phase, suggesting a higher turbulence or unresolved bulk flows. These results appear robust across a range of model assumptions, but quantitative conclusions ultimately require future improvements in the robustness of the low-energy gain calibration of the XRISM/Resolve detector.
\item The kinetic energy of the uplifted gas is estimated to be $\sim5 \times 10^{56}$ erg, almost one order of magnitude less than the corresponding gravitational potential energy. 
\end{itemize}

Interestingly, our findings contrast with predictions from some AGN feedback models (e.g., TNG), which suggest that the hottest gas should exhibit the highest velocities \citep{nelson2019} -- though note that this does not take into account the effect of projection or emission measure weighting. In M87, we observe the opposite trend, with the cooler gas indicating more pronounced dynamical signatures. This highlights the need for further observational and theoretical work to understand the coupling between AGN energy injection and multiphase gas dynamics.

\section*{Data availability}
The XRISM M87 data sets used in this work can be accessed by querying JAXA's Data ARchives and Transmission System (DARTS) \footnote{\hyperlink{}{https://darts.isas.jaxa.jp/app/query/astroquery/basic.php?satelites=xrism}}.

\begin{acknowledgements}
SRON is supported financially by The Netherlands Organisation for Scientific Research (Nederlandse Organisatie voor Wetenschappelijk Onderzoek, NWO). 
AS acknowledges the Kavli IPMU for their continued hospitality. 
This work was supported by JSPS Core-to-Core Program (grant number:JPJSCCA20220002). Contributions from M. Loewenstein are based upon work supported by NASA under award number 80GSFC24M0006. The material is based upon work supported by NASA under award number 80GSFC21M0002. BRM, ND, and MC thank the Canadian Space Agency and NSERC for support.  HRR is supported by an Anne McLaren Fellowship funded by the University of Nottingham.
\end{acknowledgements}

\bibliography{m87.bib}
\bibliographystyle{aa}

\begin{appendix}

\section{Spectral plots}\label{app:spec}

\begin{figure*}[!htb]
    \centering
\includegraphics[width=\linewidth]{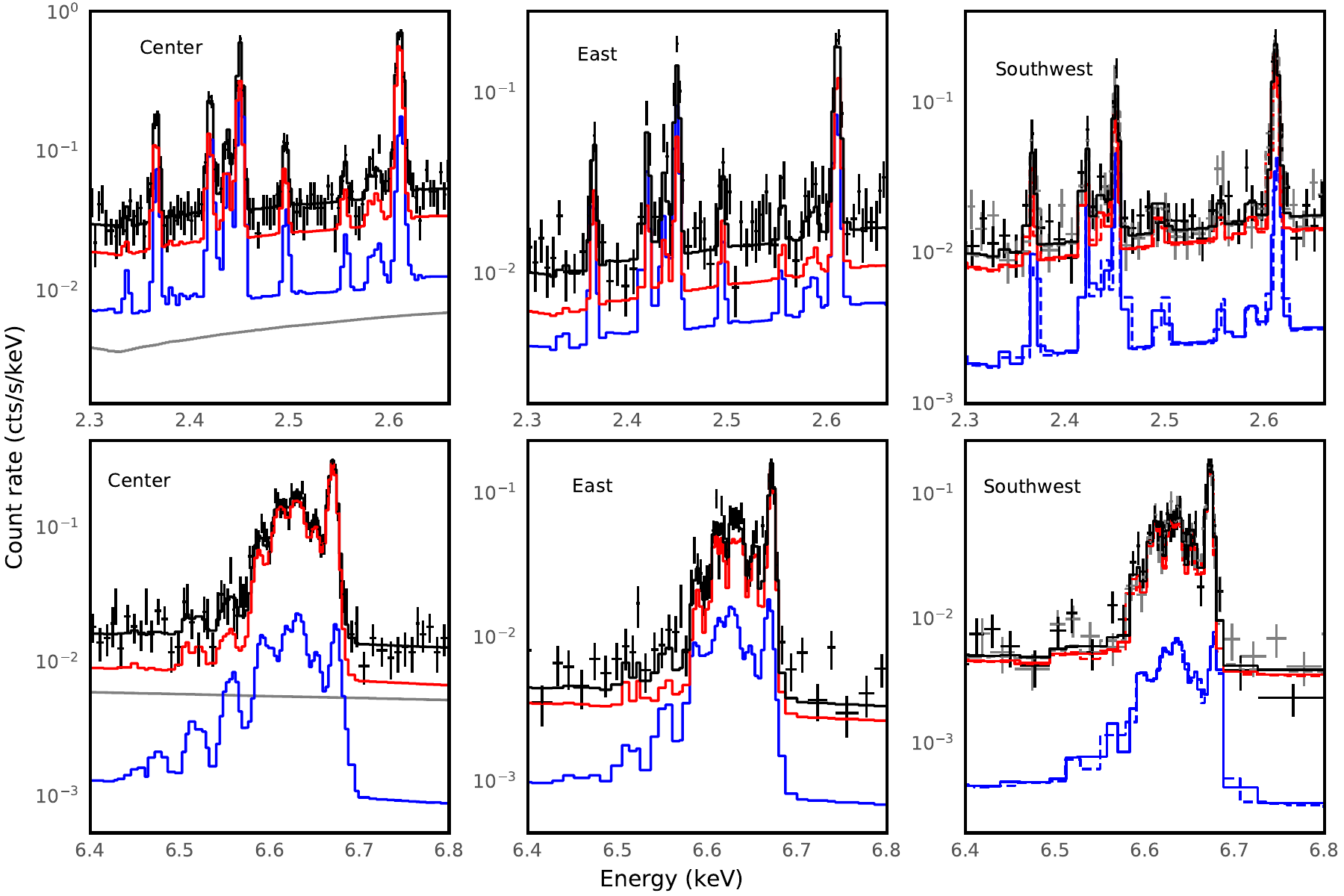}
    \caption{Comparison of the data and best fit two-temperature model for the C, E, and SW pointings. For the central pointing, we show the model described in Section \ref{sect:multiT_c}, while for the E and SW we plot `model A' as described in Section \ref{sect:joint}. For SW, results for pointing SW1 are shown with dashed lines and gray data points, while for SW2 with solid lines and black data points. We zoom in on the 2--3 keV and 6--7 keV bands showcasing the strongest emission lines. The black solid/dashed curve represents the total model, the red curve the hotter thermal component, and the blue curve the cooler thermal component. In the C pointing, the gray line represents the AGN power-law. Scattered light and LMXB components are not explicitly shown in the interest of plot clarity, but are included in the black total model curve.}
    \label{fig:ESWmultiTplots}
\end{figure*}

In Figure \ref{fig:ESWmultiTplots}, we show the best fit two temperature models, including separate curves for each thermal component, for the C, E, and SW pointings, focusing on the 2--3 and 6--7 keV energy bands. For the central pointing, we show the model described in Section \ref{sect:multiT_c} and Table \ref{tab:2T2v_fovfits}, while for the E and SW we plot `model A' as described in Section \ref{sect:joint} and Table \ref{tab:2T2v_joint}.

\section{Model uncertainties on the central pointing fits}\label{app:center}

\begin{figure*}[!htb]
    \centering
    \includegraphics[width=0.9\linewidth]{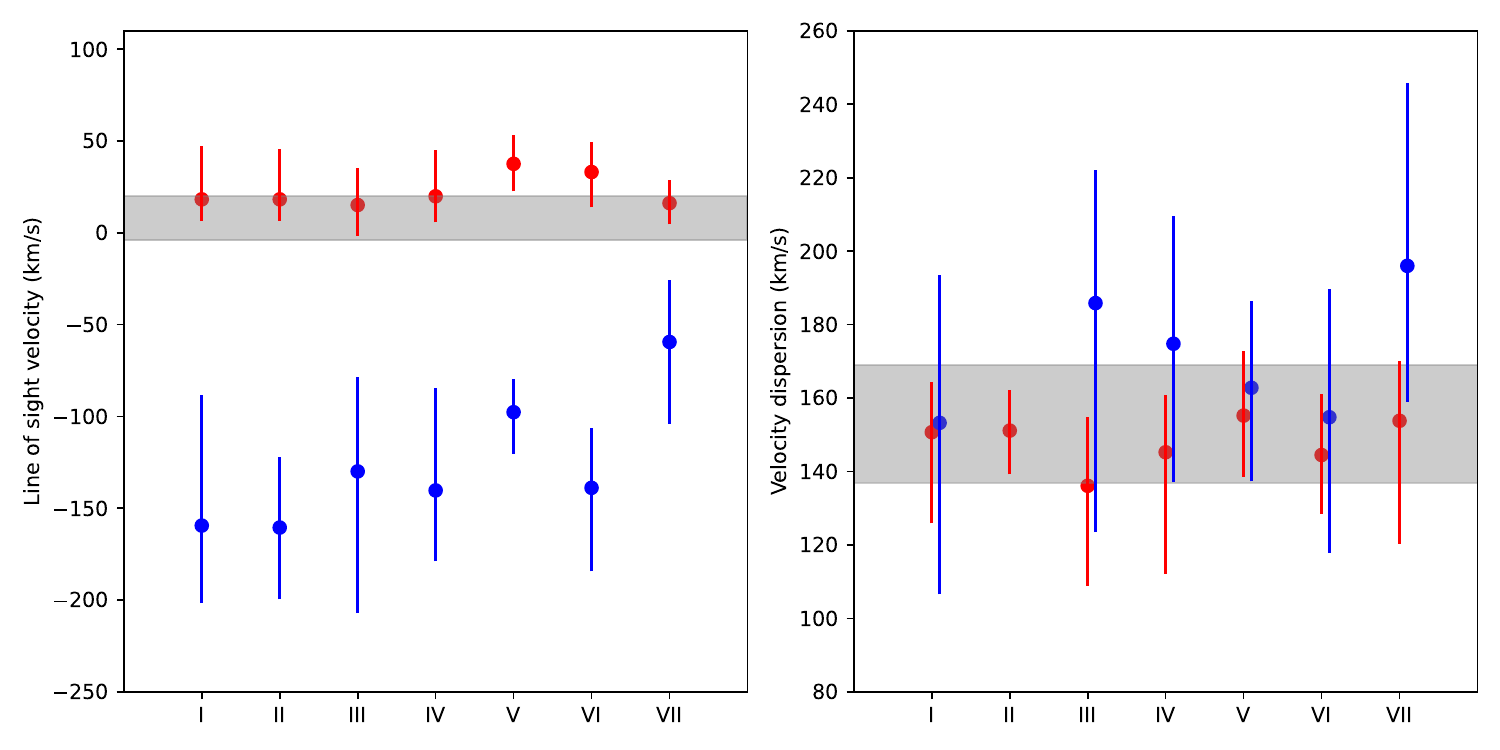}
    \caption{Multi-velocity fits to the M87 Central pointing, following the model variants described in Appendix \ref{app:center}. The best-fit velocities for the hotter component are shown in red, and for the cooler component in blue. The results from \citetalias{XRISM_M87I} are overplotted in gray. }
    \label{fig:sysC}
\end{figure*}

Here, we investigate whether the multi-phase velocity structure reported for the central pointing is sensitive to various model assumptions and uncertainties. Specifically we compare:
\begin{itemize}
    \item model I: as presented in Section \ref{sect:multiT_c}.
    \item model II: like model I, but coupling the velocity dispersions of the two gas phases (since their values are statistically consistent).
    \item model III: like model I, but using SPEXACT v3.08.01.
    \item model IV: like model I, but the abundances of Si, S, Ar, Ca, Cr, Fe, Ni for the cooler phase are decoupled from the hotter gas and coupled to each other, and left as one additional free parameter in the fit. The best fit absolute abundance of the cooler gas is only weakly constrained at 1.67(-0.41,+0.74) Solar. 
    \item model V: like model I, but fixing the LMXB flux to its lowest plausible value computed from the measured stellar mass distribution \citep{Nuker2022} and the ratio of LMXB luminosity to mass from \cite{Revnivtsev2014}. This estimate neglects any additional contribution associated with the high M87 specific frequency of Globular Clusters \citep{Chandra2018}. It amounts to a total flux in the central pointing, in the 2--7 keV band, of $10^{-12.16}$ erg/s/cm$^2$, i.e. about twice lower than the best-fit value reported in Table \ref{tab:2T2v_fovfits}.
    \item model VI: following the approach presented in \cite{XRISM2025_NGC3783cross}, we modify the nominal arf by a correction function aimed at mitigating remaining calibration uncertainties related to the gate valve thickness. This is described as: 
\begin{equation}\label{eqn:farf}
    A_{\rm eff,corr} (E) = \frac{A_{\rm eff} (E) }{1-0.00333 \times e^{(4-E)/0.582}}
\end{equation}
    \item model VII: as model I, but modifying the spectral response file to introduce a linear gain transformation, corresponding to no shift at 6.7 keV, and a -0.33 eV shift at 2.5 keV (similar to `model K' in Section \ref{sect:joint_mosys}).
\end{itemize}

The results are presented in Figure \ref{fig:sysC}. The cooler gas phase is significantly blue-shifted compared to the hotter ambient for all models, however the magnitude of this blueshift decreases considerably when the -0.33 eV shift is introduced in the \texttt{rmf}. We conclude that the velocity difference between the cooler and hotter gas phases in the central pointing is not robust to instrumental calibration errors, and do not further discuss its physical interpretation. 

\section{Impact of additional calibration uncertainties on the multi-velocity structure in the E and SW arms}\label{app:offsets}

\begin{figure*}[!htb]
       \centering
\includegraphics[width=\linewidth]{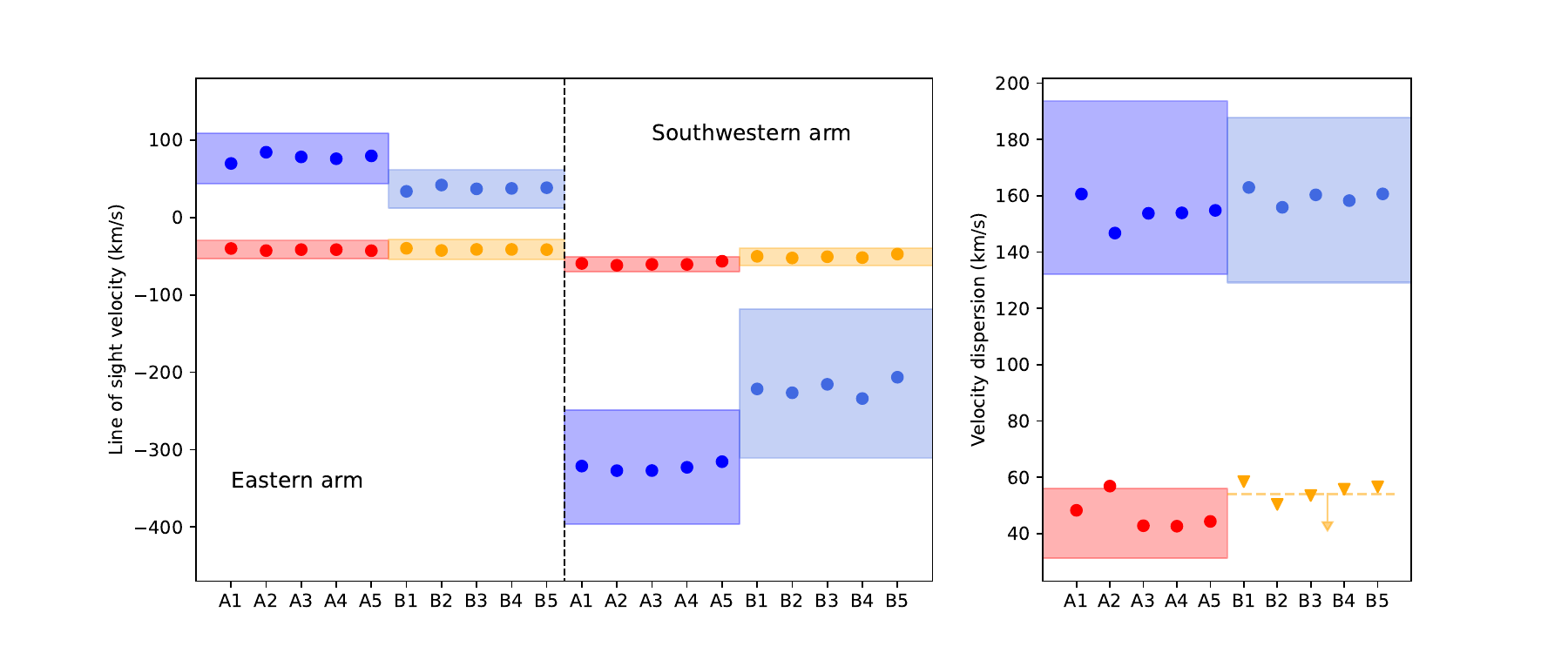}
    \caption{The impact of various calibration uncertainties on the velocities of the E and SW arms. 
    Model variants (1,2): scattering from the cluster center is decreased and increased by ±30\%. Model variants (3,4): the NXB level is decreased and increased by ±10\%. Model variant 5: the energy dependent effective area is modified according to Eq. \ref{eqn:farf}. These modifications are applied both to "model A" (for which the velocities of the hotter component are plotted in red, and for the cooler component in blue), and to "model B" (for which the results of the hotter/cooler component are shown in orange and light blue, respectively). In each case, the best-fit value is shown as a circle, or in the case a parameter was consistent with 0 we show the $2 \sigma$ upper limit as a triangle. The bands of corresponding colors show the $\pm1\sigma$ statistical uncertainty interval of the fiducial models A and B (as described in Sections \ref{sect:joint} and \ref{sect:joint_mosys}, without any systematic variations included).}
    \label{fig:ins_sys} 
\end{figure*}

In Section \ref{sect:joint_mosys}, we explored the robustness of our velocity measurements to various astrophysical model assumptions and uncertainties in the detector gain. Additional calibration uncertainties can also play a role. Here, we investigate a few other relevant parameters, namely:

\begin{itemize}
    \item the uncertainty on the amount of scattered light from the Central pointing into the offsets. To this end, we vary the normalization of the scattered thermal components by ±30\% from their best fit value.
    \item the uncertainty on the NXB level. To this end, we scale the NXB spectra by ±10\% from their original value.
    \item the calibration of the relative effective area, in particular at lower energies (below $\sim$2.6 keV), where the gate valve transmission has not been directly measured prior to launch and can be uncertain. Here, we follow the approach presented in \cite{XRISM2025_NGC3783cross} and modify the \texttt{arf} according to Eqn\ref{eqn:farf}.
\end{itemize}

We perform these checks both for our fiducial "model A", as well as for "model B" which showed the differences between the cooler gas velocity in the E and SW at the smallest (though still considerable) statistical significance. The results are shown in Figure \ref{fig:ins_sys}, and demonstrate that these additional calibration uncertainties do not affect our results. 

We further note, in passing, that the expected LMXB fluxes in the offset pointings, computed using the same method as `model V' in Appendix \ref{app:center}, range from $10^{-12.95}$ to $10^{-12.56}$ erg/s/cm$^2$. These expectations are close to the best-fit values shown in Tables \ref{tab:2T2v_fovfits} and \ref{tab:2T2v_joint}, especially when we consider that LMXB scattering from the C pointing into the offsets has not been accounted for and can contribute to increasing the observed values.

\end{appendix}

\end{document}